\documentclass[3p,times,procedia]{elsarticle}
\usepackage{nupha_ecrc}
\usepackage{hyperref}
\hypersetup{
    colorlinks=true,
    linkcolor=blue,
    filecolor=magenta,      
    urlcolor=cyan,
}
  
\volume{00}

\firstpage{1}

\journalname{Nuclear Physics A}

\runauth{G.K. Krintiras (on behalf of the CMS Collaboration}

\jid{nupha}

\jnltitlelogo{Nuclear Physics A}

\usepackage{amssymb}

\usepackage{lineno}

\usepackage{ptdr-definitions}
\usepackage{hepunits}
\usepackage{heppennames2}

\usepackage[figuresright]{rotating}

\newcommand{\rootsNN} {\ensuremath{\sqrt{\smash[b]{s_{_{\mathrm{NN}}}}}}\xspace}
\newcommand{\roots} {\ensuremath{\sqrt{\smash[b]{s}}}\xspace}
\newcommand{\PbPb}{\ensuremath{\mathrm{Pb}\mathrm{Pb}}\xspace}
\newcommand{\NN}{\ensuremath{\mathrm{N}\mathrm{N}}\xspace}
\newcommand{\pp}{\ensuremath{\Pp\Pp}\xspace}

\newcommand{\stt}{\ensuremath{\sigma_{\ttbar}}\xspace}

\newcommand{\RecoLumi}{\ensuremath{ 1.7\pm 0.1\nbinv}}
\newcommand{\empm}{\ensuremath{\Pe^\pm \Pgm^\mp}}
\newcommand{\mmpm}{\ensuremath{\Pgm^+ \Pgm^-}}
\newcommand{\empe}{\ensuremath{\Pe^+ \Pe^-}}
\newcommand{\eb}{\ensuremath{\varepsilon_{\cPqb}}}
\newcommand{\dQGP}{\ensuremath{\delta_{\text{QGP}}}}
\newcommand{\levone}{\ensuremath{2\ell_\text{OS}}}
\newcommand{\levtwo}{\ensuremath{2\ell_\text{OS}{}+{}\text{b-tags}}}
\providecommand{\PN}{\ensuremath{\mathrm{N}}\xspace}
\newcommand{\VV}{\ensuremath{\cmsSymbolFace{V}\cmsSymbolFace{V}}\xspace}
\newcommand{\mub}{\ensuremath{\,\mu\text{b}}\xspace}

\newcommand{\zvalexptwol}{\ensuremath{ 4.8}}
\newcommand{\zvalobstwol}{\ensuremath{ 3.8}}
\newcommand{\zvalexptwoljets}{\ensuremath{ 6.0}}
\newcommand{\zvalobstwoljets}{\ensuremath{ 4.0}}

\newcommand{\muobstwol}{\ensuremath{ 0.81^{+0.26}_{-0.23}}}

\newcommand{\muobstwoljets}{\ensuremath{ 0.64^{+0.22}_{-0.20}}}
\newcommand{\stttwol}{\ensuremath{2.56\pm 0.82\,\text{(tot)}}}
\newcommand{\stttwoljets}{\ensuremath{2.02\pm 0.69\,\text{(tot)}}}

\begin{document}

\begin{frontmatter}


\dochead{XXVIIIth International Conference on Ultrarelativistic Nucleus-Nucleus Collisions\\ (Quark Matter 2019)}

\title{Evidence for top quark production in nucleus-nucleus collisions}

\author[GKK]{G.K. Krintiras\fnref{colab}\fnref{fund}}
\fntext[colab]{On behalf of the CMS Collaboration}
\fntext[colab]{Supported by the Nuclear Physics program \href{https://pamspublic.science.energy.gov/WebPAMSExternal/Interface/Common/ViewPublicAbstract.aspx?rv=d1ddcae6-235b-4163-ae34-01fce58e5f90&rtc=24&PRoleId=10}{DE-SC0019389} of the U.S. Department of Energy.}
\address[GKK]{The University of Kansas}
\ead[url]{cern.ch/gkrintir}

\begin{abstract}
Droplets of quark-gluon plasma (QGP), an exotic state of strongly interacting quantum chromodynamics (QCD) matter, are routinely produced in heavy nuclei high-energy collisions. Although the experimental signatures marked a paradigm shift away from expectations of a weakly coupled QGP, a challenge remains as to how the locally deconfined state with a lifetime of a few {}\fm\ can be resolved. The only colored particle that decays mostly within the QGP is the top quark. Here we demonstrate, for the first time, that top quark decay products are identified, irrespective of whether interacting with the medium (bottom quarks) or not (leptonically decaying \PW bosons). Using $1.7 \pm 0.1\,\nbinv$ of lead-lead ($A = 208$) collision data recorded by the CMS experiment at a nucleon-nucleon center-of-mass energy of 5.02 \TeV, we report evidence of top quark pair (\ttbar) production.  Dilepton final states are selected, and the cross section (\stt) is measured from a likelihood fit to a multivariate discriminator using lepton kinematic variables. The \stt measurement is additionally performed considering the jets originating from the hadronization of bottom quarks, which improve the sensitivity to the \ttbar signal process. After background subtraction and analysis corrections, the measured \stt is \stttwol\ and \stttwoljets \mub in the two cases, respectively, consistent with predictions from perturbative QCD.
\end{abstract}


\end{frontmatter}


\section{Introduction}
\label{sec:Into}

Complementary methods are used to study an exotic form of matter, produced in heavy nuclei high-energy collisions
and referred to as the quark-gluon plasma (QGP).
The characteristic feature of experimental signatures is their sensitivity to initial- or final-state effects
integrated over the QGP lifetime, the latter increasing as a function of nucleon-nucleon (\NN) center-of-mass energy \rootsNN and the atomic mass $A$ of the ions being collided.
At variance with measurements considered so far in the literature, \eg, Ref.~\cite{Busza:2018rrf}, top quark (\PQt) offers the unique opportunity
to resolve the QGP thanks to its mean lifetime $\tau_{\PQt} \approx 0.15\,\fm$.
Although the feasibility of top quark studies with nuclear collisions was demonstrated recently~\cite{Sirunyan:2017xku},
they remained inaccessible in nucleus-nucleus collisions because of the small amount of data accumulated so
far at the CERN LHC and the lower \rootsNN available, \eg, at the BNL RHIC.
At the end of 2018, LHC was reconfigured to provide its four major experiments with lead-lead (PbPb) collisions at $\rootsNN = 5.02\,\TeV$ at an unprecedented collision rate.
The amount of data recorded by the CMS experiment~\cite{Chatrchyan:2008zzk} corresponds to an integrated luminosity of about \RecoLumi~\cite{CMS-PAS-LUM-17-002}.

At the LHC, top quark cross sections are dominated by pair production (\ttbar) via gluon-gluon ({\cPg{}\cPg}) fusion, \ie, the partonic reaction $\cPg\cPg \to \ttbar+X$, and are
computable with high accuracy in perturbative quantum chromodynamics (QCD)~\cite{Czakon:2011xx,Czakon:2013goa,Catani:2019hip}.
The top quark is thus a theoretically precise probe of the initial state and, in particular, the gluon parton distribution function (PDF)~\cite{Sirunyan:2017ule}. 
Top quarks then decay mostly within the strongly interacting medium and promptly without hadronizing into bottom (\cPqb) quarks and \PW\ bosons;
the latter further decays to leptons or quarks and defines the final state~\cite{TANABASHI:2018OCA}.
The detailed study of top quark decay products provides novel insights into the mechanisms of medium-induced parton energy loss, a final-state effect. 
On the one hand, \cPqb quarks are ideally suited to serve as a standard candle of the amount of energy suppression for \cPqb quark jets~\cite{Chatrchyan:2013exa} emerged
almost simultaneously with the heavy ion collisions~\cite{Baskakov:2015nxa}.
On the other hand, hadronically decaying \PW\ bosons are not immediately resolved by the medium, 
and hence probe the QGP density evolution at different space-time scales~\cite{Apolinario:2017sob,Dainese:2016gch}.

\section{Methods}
\label{sec:Methods}

In these proceedings, we describe the evidence for \ttbar production in nucleus-nucleus collisions~\cite{CMS-PAS-HIN-19-001}. 
We first identify \ttbar signal events based on the leptonic $\PW^{\pm} \to \ell^{\pm}\PGn$ decays, with $\ell$ being either an electron (\Pe) or muon (\Pgm).
The resulting dilepton final states (\empm, \mmpm, and \empe) present a typical signature of two energetic, oppositely charged (OS),
and isolated from nearby hadronic activity leptons (``\levone''), momentum imbalance from the undetected neutrinos (\PGn), and two \cPqb\ jets.
Dilepton final states are of particular interest since leptons propagate through the produced medium regardless of its nature,
thereby providing favorable conditions for the detection of \ttbar production.
Since the feasibility to reconstruct \cPqb quarks out of the QGP may be impacted by the sizeable suppression evidenced in data, 
the ``\levtwo'' method further relies on a data-based estimate for the \cPqb\ jet identification (``tagging'') performance, measured in terms of the \cPqb\ jet identification efficiency \eb.
Finally, we extract the \ttbar cross section (\stt) from a combined maximum-likelihood fit to a multivariate discriminator using lepton kinematic variables, 
independently with the \levone\ and \levtwo\ methods.

The $\PN\PN \to \ttbar+X$ process ($\PN = \Pp,\Pn$) is simulated at next-to-leading order (NLO) using the EPPS16 NLO nuclear PDF~\cite{ESKOLA:2016OHT}.
Sources of the dominant background, \ie, Drell--Yan production of quark-antiquark annihilation into lepton-antilepton pairs 
through \cPZ boson or virtual-photon exchange (referred to as ``$\cPZ{}/\gamma^{*}$'')
and nonprompt leptons, \eg, \PW boson production with additional jets (\PW{}+jets) are generated at NLO; both are further corrected with scaling factors derived from data.
Subdominant contributions from single top quark plus \PW{} boson events ($\PQt\PW$) and \PW{}\PW, \PW{}\cPZ, and \cPZ{}\cPZ\ production (collectively referred to as ``\VV'') are simulated at NLO too. 

Gradient boosted decision trees (BDTs) are set up to maximally discriminate genuine leptons with high transverse momentum (\pt) between the \ttbar signal and background processes.
The BDTs exploit kinematic properties of the leading- and subleading-\pt leptons (referred to as ``$\ell_1$'' and ``$\ell_2$'', respectively).
We perform a maximum-likelihood fit to binned BDT distributions in the \empm, \mmpm, and \empe\ final states, accounting for all sources of uncertainty, \ie,
statistical and systematic, and their correlations. The simultaneous analysis of these events allows us measuring the \ttbar signal strength $\mu$, \ie,
the ratio of the observed \stt\ to the expectation from theory. The best fit value of $\mu$ and its uncertainty $\Delta\mu$ (corresponding to a 68\% confidence level) are extracted independently with the \levone\ and \levtwo\ methods. For the latter, we correlate the number of \ttbar signal events in the \cPqb{}-tagged jet categories based on multinomial probabilities,
using \eb\ and a parameter (\dQGP) accounting for medium-induced suppression of \eb. We allow \eb\ to receive different values for the two \cPqb\ jets, i.e.,
$\eb \to \eb^*=(1-\dQGP) \times \eb$, motivated by a path-length dependence of the parton energy loss.

\section{Results}
\label{sec:Results}

\begin{figure}[!htp]
\centering
\includegraphics[width=0.32\textwidth]{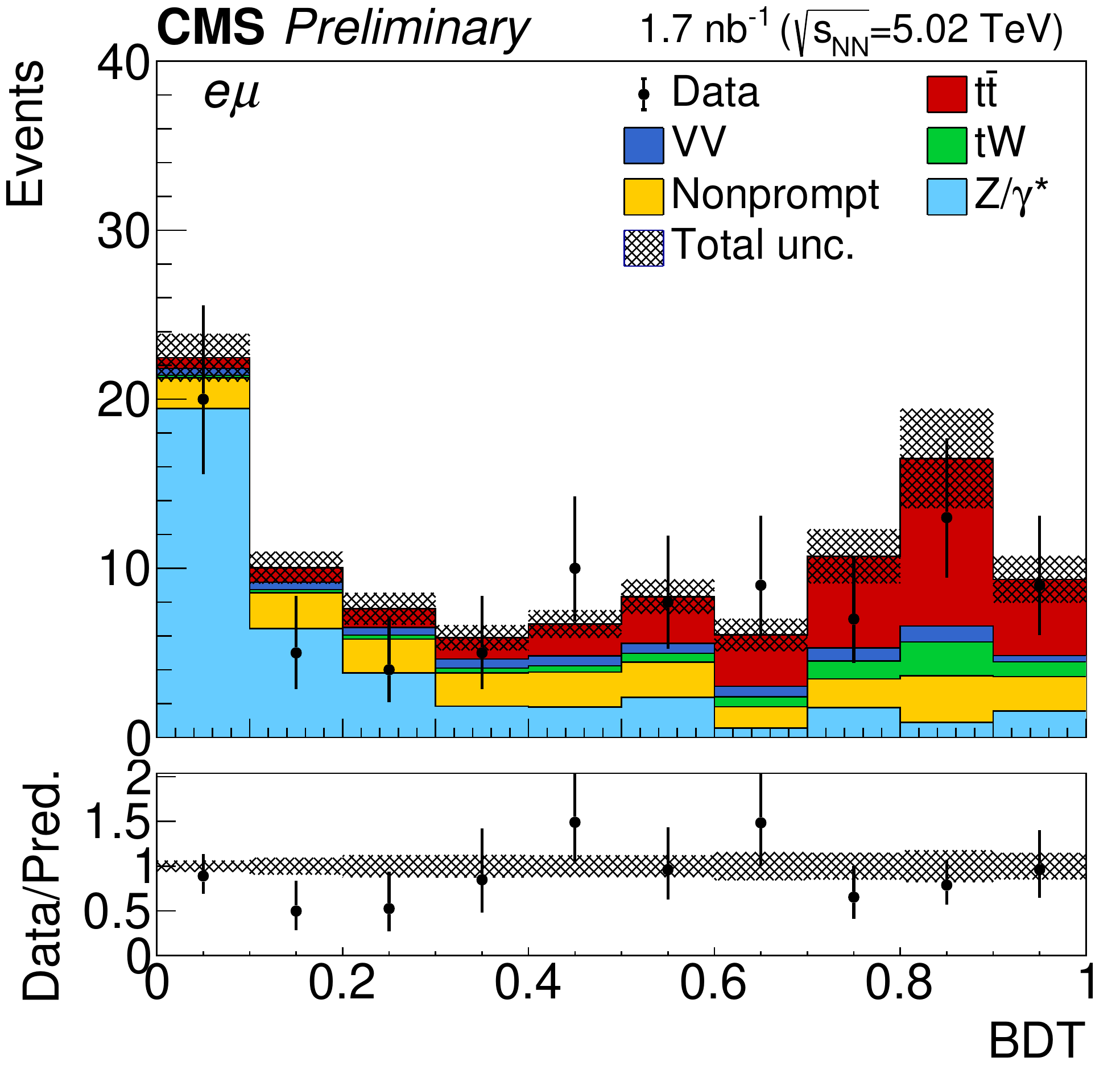}
\includegraphics[width=0.32\textwidth]{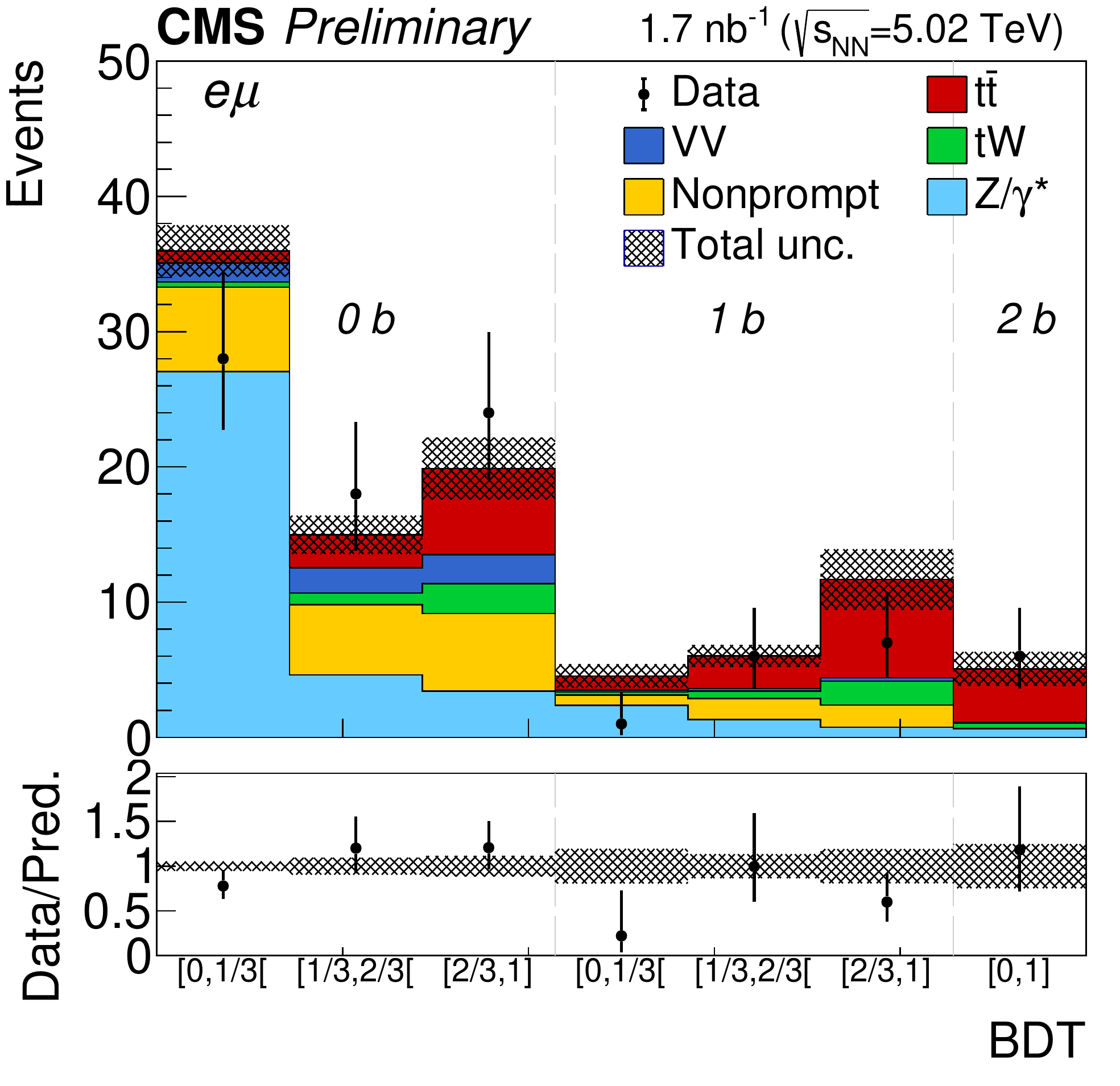}
\includegraphics[width=0.265\textwidth]{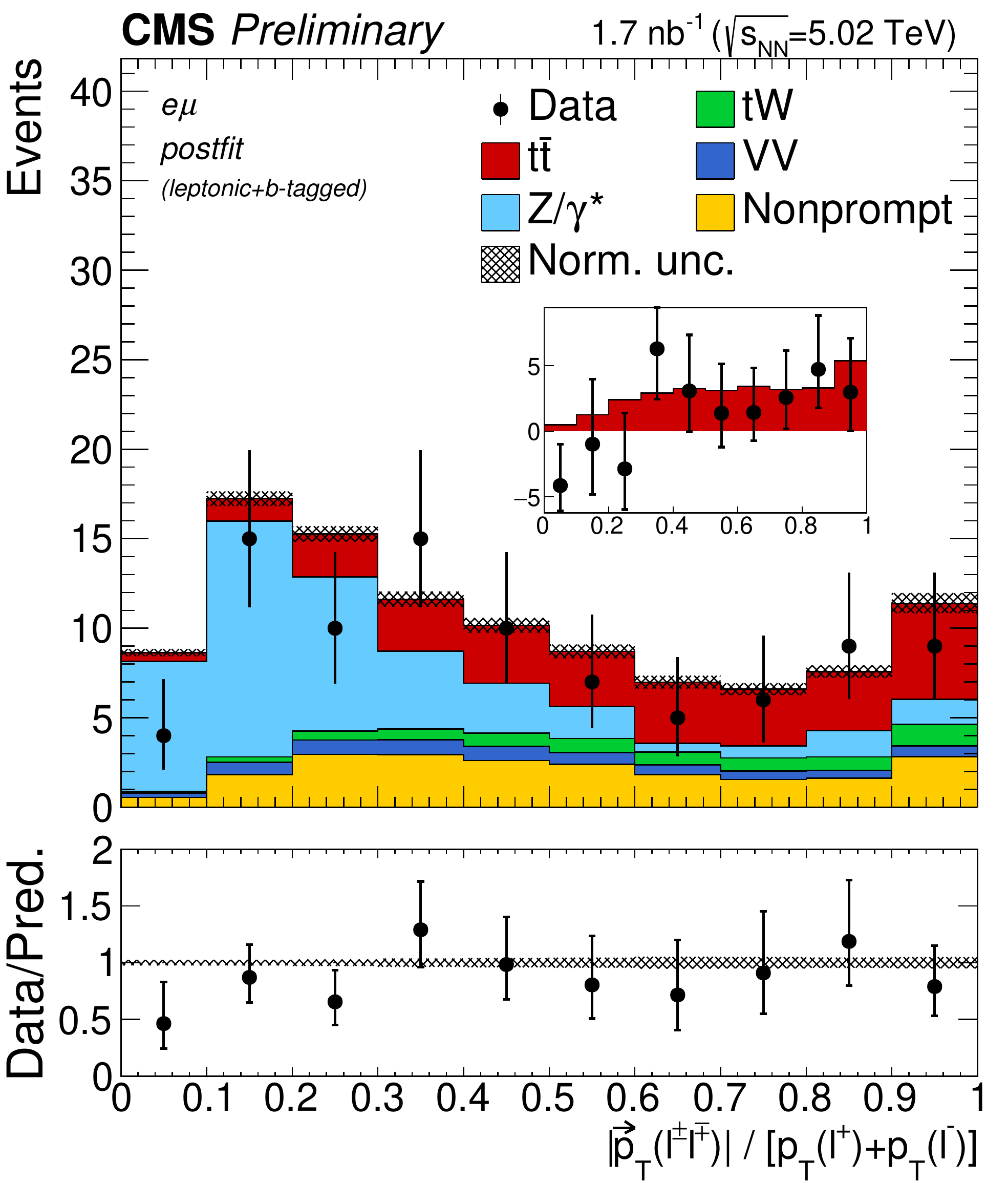}
\caption{
  Postfit expected (histograms) and observed (points) binned BDT (left, middle) and sphericity (right) distributions in the \empm\ final state with the \levone\ (left, right) and \levtwo\ (middle) methods.
  The comparison between the \ttbar\ signal and the background-subtracted data is shown for the postfit distributions as inset panel (right).
  The vertical bars on the points represent the statistical uncertainty in data (68\% Clopper-Pearson intervals).
  The hatched regions show the postfit uncertainty in the sum of \ttbar\ signal and background. 
  The lower panels display the ratio of the observed data to the predictions, including the \ttbar signal, with bars and bands, respectively,
  representing the statistical and total uncertainties in the prediction~\cite{CMS-PAS-HIN-19-001}.
    }
\label{fig:results}
\end{figure}

The postfit expected and observed BDT distributions are shown in Fig.~\ref{fig:results} in the \ttbar{}-enriched \empm\ final state with the \levone\ ($\mu=\muobstwol$, left) and \levtwo\ ($\mu=\muobstwoljets$, middle) methods. 
We found the classifier to separate well the \ttbar signal from the \cPZ/$\gamma^{*}$ background in the \mmpm\ and \empe\ final states, 
enhancing our confidence about its applicability to the \empm\ final state. The inclusive \stt\ is then obtained multiplying 
the best fit $\mu$ value by the theoretical expectation. Accounting for the acceptance corrections, we measure
\stt to be \stttwol\ and \stttwoljets \mub in the combined \empm, \mmpm, and \empe\ final states, with a relative total uncertainty of 32 and 34\% with the \levone\ and \levtwo\ methods, respectively.

The compatibility of the data with the background-only hypothesis is evaluated using
a profile-likelihood ratio as a test statistic, including all sources of systematic uncertainty.
The probability for the background to mimic an excess of events as large or larger than that observed in data is measured using $p$ values. 
The latter values are then expressed in terms of Gaussian tail probabilities, given in units of standard deviation ($\sigma$).
The background-only hypothesis is excluded with statistical significance of \zvalobstwol\ (\zvalexptwol) and \zvalobstwoljets\ (\zvalexptwoljets)\,$\sigma$
with the \levone\ and \levtwo\ methods, respectively. The $p$ value of the difference between observed and expected significance, given in parentheses, is 0.18 and 0.05, respectively.
To further examine the hypothesis that the selected data are consistent with the production of top quarks, 
we study the sphericity $s=\pt(\ell^\pm \ell^\mp)/\left( \pt (\ell_1) + \pt (\ell_2) \right)$, a global-event property. 
We found $s$ (Fig.~\ref{fig:results}, right) to perform well in terms of separating the \ttbar signal ($s\sim1$) from the background ($s\sim0$)
and to exclude the background-only hypothesis at a similar level.

\begin{figure}[!htb]
\centering
\includegraphics[width=0.50\textwidth]{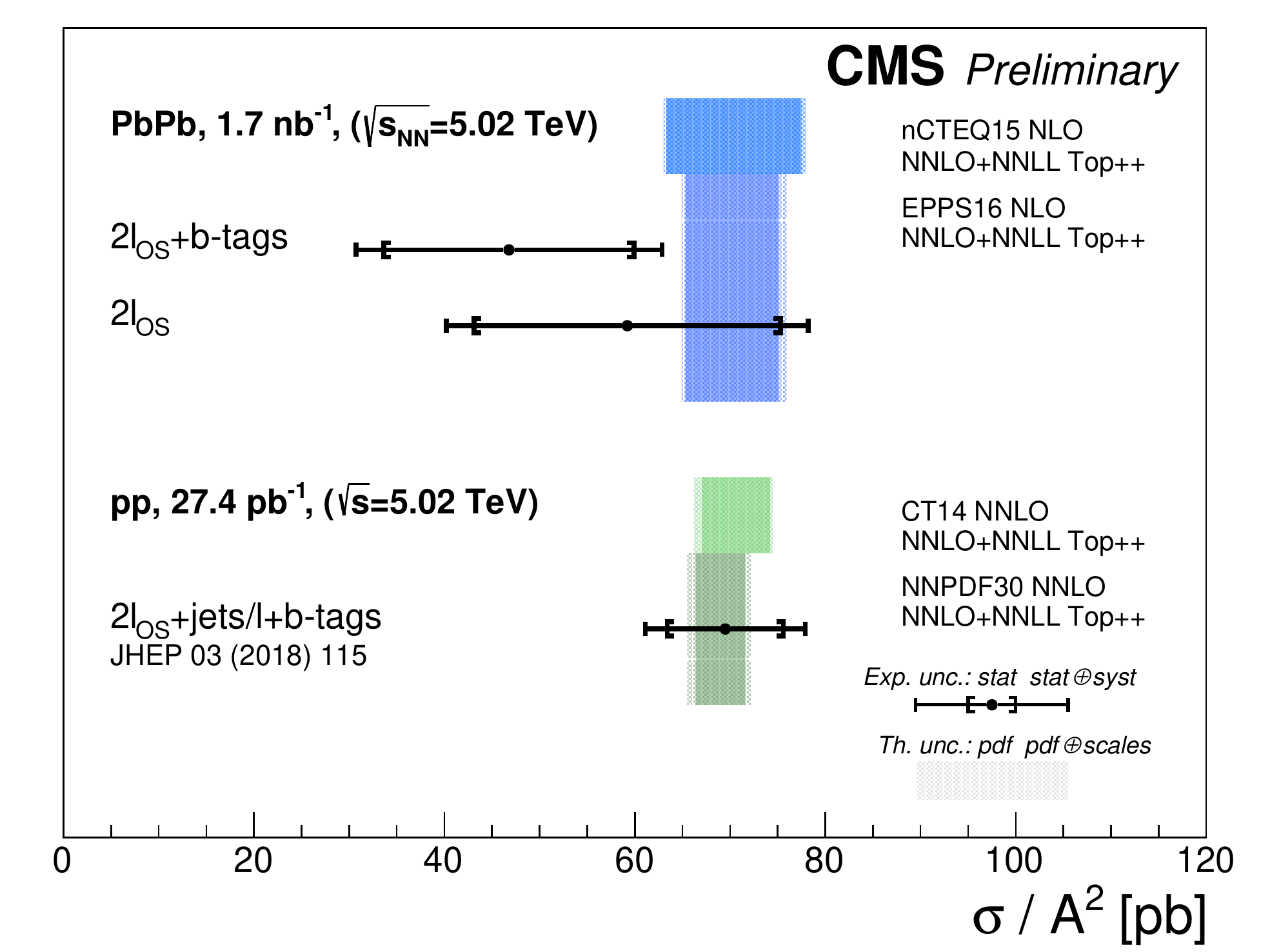}
\caption{The inclusive \ttbar\ cross section measured in the combined \empm, \mmpm, and \empe final states in \PbPb\ collisions (divided by the mass number squared, $A^2$),
  compared to predictions~\cite{Czakon:2013goa}, and \pp\ results at $\rootsNN=5.02$ \TeV~\cite{Sirunyan:2017ule}.
  The total experimental error bars (theoretical error bands) include statistical and systematic (PDF and QCD scale) uncertainties added in quadrature~\cite{CMS-PAS-HIN-19-001}.
  }
\label{fig:summary}
\end{figure}

Figure~\ref{fig:summary} presents the extracted \stt, including the measurement at $\roots=5.02$ \TeV~\cite{Sirunyan:2017ule} in proton-proton (\pp) collisions,
and compared to state-of-the-art theoretical QCD predictions~\cite{Czakon:2013goa}.

\section{Summary}
\label{sec:Summary}

In summary, the top pair production cross section is measured for the first time in nucleus-nucleus collisions, 
using lead-lead data at $\rootsNN = 5.02\,\TeV$ with an integrated luminosity of \RecoLumi.
The measurement is performed analyzing separately events containing at least one pair of oppositely charged leptons (electrons or muons) and 
jets originating from the hadronization of bottom quarks.
The observed statistical significance of the \ttbar\ signal against the background-only hypothesis is \zvalobstwol\ and \zvalobstwoljets\ standard deviations in the two cases, respectively.
The measured cross section is \stttwol\ and \stttwoljets \mub, respectively, consistent with the expectations from scaled proton-proton data as well as perturbative quantum chromodynamics calculations. 
The measurement paves the way for detailed investigations of top quark production in nuclear interactions with increased LHC heavy ion luminosities or future higher-energy colliders. 
In particular, it provides a novel tool for probing nuclear parton distribution functions, the mechanisms of parton energy loss, and the medium opacity at different space-time scales.

\bibliographystyle{elsarticle-num}
\bibliography{nupha-template}

\end{document}